\documentclass[twocolumn]{aastex62}
\usepackage[utf8]{inputenc}
\usepackage{amssymb}
\usepackage{amsmath}
\usepackage{acronym}
\usepackage{hyperref}
\usepackage{xcolor}

\usepackage{mathtools} 
\makeatletter
\newcommand{\raisemath}[1]{\mathpalette{\raisem@th{#1}}}
\newcommand{\raisem@th}[3]{\raisebox{#1}{$#2#3$}}
\makeatother
\NewDocumentCommand{\newbbar}{O{0pt} O{0pt}}{
  \ensuremath{\mathrlap{\raisemath{#2}{\hspace*{#1}{\mathchar'26\mkern-9mu}}}b}%
}

\newcommand{\msun}{\ensuremath{M_\odot}}

\acrodef{GW}[GW]{gravitational-wave}
\acrodef{BBH}[BBH]{Binary black hole}  
\acrodef{PE}[PE]{parameter estimation}
\acrodef{KDE}[KDE]{kernel density estimation}

\usepackage[normalem]{ulem} 
\usepackage{soul}

\begin{document}

\title{Binary vision: The merging black hole binary mass distribution via iterative density estimation}

\correspondingauthor{Jam Sadiq}
\email{jsadiq@sissa.it}  
\author[0000-0001-5931-3624]{Jam Sadiq}
\affil{Instituto Galego de F\'{i}sica de Altas Enerx\'{i}as, Universidade de Santiago de Compostela, Santiago de Compostela, Galicia, Spain}
\affil{SISSA, Via Bonomea 265, 34136 Trieste, Italy and INFN Sezione di Trieste}
\author[0000-0003-1354-7809]{Thomas Dent}
\affil{Instituto Galego de F\'{i}sica de Altas Enerx\'{i}as, Universidade de Santiago de Compostela, Santiago de Compostela, Galicia, Spain}
\author[0000-0002-9716-1868]{Mark Gieles}
\affil{ICREA, Pg. Lu\'{i}s Companys 23, E08010 Barcelona, Spain}
\affil{Institut de Ci\`{e}ncies del Cosmos (ICCUB), Universitat de Barcelona (IEEC-UB), Mart\'{i} Franqu\`{e}s 1, E08028 Barcelona, Spain}

\date{\today}

\begin{abstract} 
  \ac{BBH} systems detected via \ac{GW} emission are a recently opened astrophysical frontier with many unknowns and uncertainties.  Accurate reconstruction of the binary distribution with as few assumptions as possible is desirable for inference on formation channels and environments.  Most population analyses have, though, assumed a power law 
  in binary mass ratio $q$, and/or assumed a universal $q$ distribution regardless of primary mass.  
  Kernel density estimation (KDE)-based methods allow us to dispense with such assumptions and directly estimate the joint binary mass distribution. We deploy a self-consistent iterative method to estimate this full \ac{BBH} mass distribution, finding local maxima in primary mass consistent with previous investigations and a secondary mass distribution with a partly independent structure, inconsistent both with a power law and with a constant function of $q$.  We find a weaker preference for near-equal mass binaries than in most previous investigations; instead, the secondary mass has its own ``spectral lines'' at slightly lower values than the primary, and we observe an anti-correlation between primary and secondary masses around the $\sim\!10\,\msun$ peak. 

\end{abstract}


\section{Introduction}
Ever since the first \ac{GW} detection revealed a binary black hole source with the---previously unsuspected---component masses of around 35\,\msun\ \citep{LIGOScientific:2016aoc,LIGOScientific:2016vpg}, 
LIGO-Virgo-KAGRA  
observations of compact binaries have continued to yield surprises, of which the binary mass distribution arguably contains the most information bearing on formation environments and channels.  In the first three observing runs of Advanced LIGO~\citep{LIGOScientific:2014pky} and Advanced Virgo~\citep{VIRGO:2014yos} the better part of 100 detections of binary compact object mergers via gravitational wave (GW) emission were made, as catalogued in the GWTC releases~\citep{LIGOScientific:2018mvr,LIGOScientific:2020ibl,LIGOScientific:2021djp,LIGOScientific:2021usb}.  Having a set of detected events it is possible to study population properties of these compact binaries and eventually draw implications from these properties on binary astrophysical formation and evolution.  Detailed investigations of the population properties of \ac{BBH} mergers, the most commonly detected source type, were undertaken in \citet{Abbott:2020gyp,KAGRA:2021duu}, focusing on several population characteristics including their component masses and spins and possible dependence on redshift. 

Among parameters estimated and studied in connection with the population properties of binary compact objects, the component masses are obtained with least uncertainty. Many parameterized and semi- or non-parametric models have been proposed to study the mass-dependence of the compact binary merger rate or the mass distribution of the merger population~\citep[see][and references therein]{KAGRA:2021duu}.  In parametric models, Bayesian hierarchical techniques are used to infer model hyper-parameter posteriors, and thus the population distribution~\citep[e.g.][]{Mandel:2009pc,Thrane:2018qnx}.  On the other hand, non-parametric models are data driven methods which learn population properties either without requiring any specific functional form, or (for semi-parametric models) allowing for generalised deviations from a given parametric model~\citep{Powell:2019nmw,Tiwari:2020otp,Tiwari:2020vym,Tiwari:2021yvr,Veske:2021qis,Rinaldi:2021bhm,Edelman:2021zkw,Edelman:2022ydv,Callister:2023tgi,Ray:2023upk,Toubiana:2023egi}.

In \cite{Sadiq:2021fin} we introduced a fast and flexible adaptive width kernel density estimation (awKDE) as a non-parametric estimation method for population reconstructions of binary black hole distribution from observed gravitational wave data.  A limitation of this method arose from the measurement uncertainty in each individual event's parameters.  Given the relatively low signal-to-noise ratios of typical detections, the binary component masses have significant uncertainties~\citep[e.g.][]{Veitch:2014wba} that can bias the overall population distribution if not properly accounted for. Specifically, for typical binary components of order $35\,\msun$ and above, detector noise results in $\lesssim$10\% mass uncertainties~\citep[e.g.][]{LIGOScientific:2021usb}; for lower mass \ac{BBH}, the chirp mass $\mathcal{M} = (m_1m_2)^{3/5}(m_1+m_2)^{-1/5}$ is measured more precisely, but the mass ratio and component masses are in general more uncertain than for heavier binaries.  These uncertainties are quantified by Bayesian \ac{PE} techniques that generate $\sim$thousands of samples for each detection, representing the posterior probability distribution over $m_1$, $m_2$ given an uninformative (flat) prior.

 In \cite{Sadiq:2021fin}, we incorporated such uncertainties either by using the median \ac{PE} sample mass as a point estimate, or by randomizing over samples.  However, these procedures inevitably broaden or over-disperse any rapid variations in the population distribution, yielding a population estimate that is biased towards being too ``smooth'' or slowly-varying.  In addition, the use of an uninformative prior introduces a bias in mass measurements as compared to a prior informed by our knowledge of the population distribution~\citep[e.g.][Appendix E]{KAGRA:2021duu}.

In this work we describe a new method to treat source property measurement uncertainties and substantially reduce such biases in the estimated population distribution.  We propose an iterative scheme  for re-weighting the \ac{PE} samples of each observed event based on a current population estimate similar in spirit to the standard expectation-maximization algorithm \citep{10.2307/2984875}. We demonstrate that this re-weighting significantly reduces biases in the population density estimate, although some bias remains, as is unavoidable for a \ac{KDE} with a relatively small number of observations from an unknown true distribution.

As an application of this new method, we estimate the full 2-dimensional component mass distribution, without assumptions on its functional form aside from the use of Gaussian kernels.  While attention has often focused on the primary component mass or on the more precisely measured chirp mass \citep[see among others][]{Dominik:2014yma,Tiwari:2020otp,Tiwari:2021yvr,Tiwari:2023xff,Edelman:2022ydv,Schneider:2023mxe,Farah:2023vsc},
less attention has been paid to the full binary distribution, either considered via the secondary $m_2$ or mass ratio $q\equiv m_2/m_1$.  We expect these parameters to bear traces of the possible BBH formation channels~\citep{Kovetz:2016kpi}, in that for dynamical (cluster) formation the two masses may be independent variates, up to a factor modelling probability of binary formation and merger~\citep{Fishbach:2019bbm,2023MNRAS.522..466A,Farah:2023swu} that typically favors near-equal masses~\citep[e.g.][]{2016PhRvD..93h4029R,OLeary:2016ayz}.  Conversely, for isolated binary evolution, some nontrivial though probably highly model-dependent correlation of component masses may arise~\citep[e.g.][]{2022ApJ...940..184V}. 

Typically, parametric models have assumed power-law $m_2$ dependence at fixed $m_1$~\citep{Kovetz:2016kpi,Fishbach:2017zga,Talbot:2018cva,LIGOScientific:2018jsj}, recovering mildly positive powers indicating a preference for equal masses.  A more detailed study using GWTC-1 events indicated a preference for the two BHs of a given binary to be of comparable mass~\citep{Fishbach:2019bbm}.  More recent non-parametric or semi-parametric studies have relaxed these assumptions, either through allowing the power-law index to vary over chirp mass~\citep{Tiwari:2021yvr}, or allowing $p(q)$ to be a free (data-driven) function~\citep{Edelman:2022ydv,Callister:2023tgi} though enforcing the same dependence over all primary masses.  \cite{Tiwari:2023xff} introduced a more flexible approach with $p(q)$ modelled by a truncated Gaussian whose parameters depend on chirp mass, finding some significant variation.  \cite{Ray:2023upk} measured the full 2-d distribution with a binned (piecewise-constant) model over $m_1$, $m_2$ (including possible redshift dependence), although they did not consider the $q$ distribution. 

Note that the mass ratio distribution presents nontrivial technical issues since (at least for lower mass \ac{BBH}) typical event measurement uncertainties are both large, and correlated with the BH orbit-aligned spin components~\citep{Cutler:1994ys,Baird:2012cu}. 
As the iterative reweighting scheme is designed to address such uncertainties, we expect it to yield a more accurate reconstruction of the full mass distribution than our previous KDE method. 

The remainder of the paper is organized as follows: in section 2 we motivate and explain our method and demonstrate it using simple one- and two-dimensional mock data.  In section 3 we apply our method to detected \ac{BBH} in GWTC-3; we compare the resulting primary mass distribution with our previous studies and extend its application to the full two-dimensional mass plane.  In section 4 we discuss the implications of our results and consider extensions of the method.  

\section{Method}

\subsection{Statistical framework}

Our general approach to population inference can be considered as similar to maximum likelihood, with uncertainties quantified via empirical bootstrap methods~\citep{bootstrapcitation}.  Given a set of observed events, if we neglect measurement uncertainty in each event's parameters, our population 
estimate is a KDE where the kernel bandwidth for each event is adjusted \citep{Breiman1977,Abramson1982,TerrellScott,SainScott} 
using an adaptive scheme to maximize the cross-validated likelihood~\citep{Sadiq:2021fin}. The adaptive bandwidth KDE~\citep{wang2011bandwidth} computes a density estimate $\hat{f}$ from observations $X_i$, $i = 1 \ldots n$ via
\begin{equation}\label{awkde_initial}
    \hat{f}(x) = n^{-1} \sum_{i=1}^{n} \frac{1}{h \lambda_i} K \left( \frac{x- X_i}{h \lambda_i} \right),
\end{equation}
where $K(\cdot)$ is the standard Gaussian kernel,
\begin{equation}
    K(z) = \frac{1}{\sqrt{2\pi}} \exp\left(\frac{-z^2}{2}\right),
\end{equation}
$n$ is the total number of samples and 
$h \lambda_i$ takes the role of a local bandwidth, with $h$ being the global bandwidth.

The local bandwidths are determined by first computing a pilot estimate $\hat{f}_0$ setting $\lambda_i = 1$ for all $i$, a standard fixed bandwidth KDE; based on this pilot density, 
we then set 
\begin{equation}
    \lambda_i  = \left( \frac{\hat{f}_0(X_i)}{g} \right)^{-\alpha},
\end{equation}
where $\alpha$ is the local bandwidth sensitivity parameter ($0 < \alpha \leq 1$) and $g$ is a normalization factor
\begin{equation}
    \log g = n^{-1} \sum_{i=1}^{n} \log \hat{f}_0(X_i).
\end{equation}
Finally the adaptive KDE $\hat{f}(x)$ is obtained by evaluating \eqref{awkde_initial} with the variable (local) bandwidth $h\lambda_i$.  
The equations are written for the case of one-dimensional data $X_i$, but the method may be applied in more dimensions by linearly transforming the data to have zero mean and unit covariance along each dimension and using an $N$-dimensional unit Gaussian kernel scaled by $h\lambda_i$.

The \ac{KDE} hyperparameters, global bandwidth $h$ and sensitivity parameter $\alpha$ 
are determined by grid search using maximum likelihood as a figure of merit.  A na\"ive ``maximum likelihood'' KDE is not well defined, as the likelihood increases indefinitely in the limit of small bandwidth kernels centered on the observations, i.e.\ delta functions~\citep[][section 2.8]{Silverman1986}. We prevent this collapse and address the bias-variance tradeoff via \textit{leave-one-out cross-validation}~\citep[][section 3.4.4]{Silverman1986} \citep[see also][]{hastie01statisticallearning}. The cross-validated (log) likelihood 
is
\begin{equation}
  \log \mathcal{L}_{\rm LOO} = \sum_{i=1}^{n} \log \hat{f}_{{\rm LOO},i}(X_i),
\end{equation}
where $\hat{f}_{{\rm LOO},i}$ is the KDE constructed from all samples \emph{except} $X_i$.  Being linear in the logarithm of the estimate at observed values, this choice will penalize \emph{relative} errors. Since we wish to obtain an accurate estimate of densities over a large dynamic range, the log likelihood is more suitable than considering absolute error or squared absolute error.

We then quantify counting uncertainties for the underlying inhomogeneous Poisson process using generalized bootstrap resampling~\citep{google_bootstrap}: we take a number of \ac{PE} samples from each detected event that is a random variable distributed as Poisson(1). 

The new aspect of this work concerns the choice of mass sample values to treat as \ac{KDE} input data $X_i$. We noted in~\cite{Sadiq:2021fin} that for nontrivial uncertainties in component masses this method, in addition to possible intrinsic biases due to the choice of a Gaussian KDE, will be biased towards an over-dispersed estimate of the true distribution.  This is expected because, with the uninformed priors used in \ac{PE}, the posterior over mass for any given event will be randomly displaced relative to the (unknown) true mass due to detector noise. Here we motivate and present our strategy for correcting this over-dispersion bias.  

Our motivation is linked to Bayesian hierarchical population inference \citep{Mandel:2009pc}, where measurement errors are treated by considering the true event properties $\vec{\theta}_i$, for detected events labelled $i=1,\ldots,N$ as nuisance parameters.  Here, the likelihood of a set of gravitational-wave data segments $d_i$ corresponding to the events is
\begin{equation} \label{eq:hier_like}
 P_N(\{d\}|\vec{\lambda}) = \prod_{i=1}^N \int P(d_i|\vec{\theta}_i) p_{\rm pop}(\vec{\theta}_i|\vec{\lambda}) \, d\vec{\theta}_i, 
\end{equation}
where $P(d_i|\ldots)$ is the likelihood for a single data segment and $p_{\rm pop}(\vec{\theta}_i|\vec{\lambda})$ is the population distribution over $\vec{\theta}_i$ for population model hyperparameters $\vec{\lambda}$ (here, for simplicity we omit selection effects).  Inference is implemented using parameter estimation (PE) samples which were generated using a standard or fiducial prior $\pi_{\rm PE}(\vec{\theta})$, often chosen as uniform over parameters of interest \citep[see e.g.][]{Veitch:2014wba,Thrane:2018qnx}.  Samples (labelled by $k$) are distributed as the posterior density $p(\vec{\theta}_i|\ldots)$ using this prior, hence
\begin{equation}
  \vec{\theta}_i^k \sim p(\vec{\theta}_i|d_i,\pi_{\rm PE}(\vec{\theta})) 
  \propto P(d_i|\vec{\theta}_i) p_{\rm pop}(\vec{\theta}_i|\vec{\lambda})
  \frac{\pi_{\rm PE}(\vec{\theta})}{p_{\rm pop}(\vec{\theta}_i|\vec{\lambda})}. 
\end{equation}
Hence, integrals over $\vec{\theta}_i$, as in Eq.~\eqref{eq:hier_like} may be performed (up to a constant factor) by summing over samples $\vec{\theta}_i^k$ \emph{re-weighted} by the ratio of the population distribution to the PE prior, $p_{\rm pop}(\vec{\theta}_i|\vec{\lambda})/\pi_{\rm PE}(\vec{\theta})$. 

Here, while not making use of this hierarchical likelihood, we remark that 
 such re-weighted \ac{PE} sample sets give an unbiased estimate of event properties if $p_{\rm pop}(\vec{\theta}_i)$ is equal to the true population distribution; conversely, the \emph{unweighted}
\ac{PE} samples give a biased estimate if $\pi_{\rm PE}(\vec{\theta})$ differs from the true population distribution.  
This observation is the basic motivation for our improved method.

A \ac{KDE} trained on points drawn randomly from \ac{PE} samples will be biased because these samples are themselves biased by the ``uninformative'' prior.  If we have access to an estimated population distribution $\hat{p}_{\rm pop}(\vec{\theta})$ that is closer to the true distribution than the PE prior is, we will obtain more accurate estimates of event properties by drawing samples weighted proportional to $\hat{p}_{\rm pop}(\vec{\theta})/\pi_{\rm PE}(\vec{\theta})$, as detailed below. 
%
The better an estimate of the true distribution we are able to obtain, the smaller will be the bias in event parameters using reweighted PE samples, and ultimately the smaller will be the bias of the KDE. 

\subsection{Iterative Reweighting}
\label{ss:iterkde}
The above discussion suggests an iterative procedure where, beginning with both biased PE samples and a biased population KDE, one may be improved in turn using the other, finally reaching a stationary state where -- ideally -- both the sample draws and the corresponding population estimates are unbiased.
This iterative strategy is similar to the expectation-maximization algorithm~\citep{10.2307/2984875}, a popular method to estimate parameters for statistical models when there are missing or incomplete data.  

Our basic algorithm follows these steps: 
\begin{enumerate}
\item For each GW event, draw Poisson distributed (with mean 1) PE samples weighted by the current estimate of population density $\hat{p}_{\rm pop}$
\item Create an awKDE from this sample set, optimizing the global bandwidth (and sensitivity parameter $\alpha$, if not fixed)
\item Update the current population estimate using one or more KDEs and the selection function, and go to step 1. 
\end{enumerate}
In more detail, in step 1 we draw PE samples with probability proportional to the ratio of $\hat{p}_{\rm pop}(\vec{\theta}_i^k)$ to the PE prior distribution.  Step 2 reproduces our previous awKDE method.  Step 3 relates the KDE of \emph{detected} events to an estimate of the true population distribution, hence in general it requires us to compensate for the selection function over the event parameter space: i.e.\ we estimate the true distribution by the KDE of detected events divided by the probability of detection, as detailed in \citep[section 3.1]{Sadiq:2021fin}. 

In step 3 we may choose to derive the updated population density $\hat{p}_{\rm pop}(\vec{\theta})$ from only the most recently calculated KDE: then the iterative process is a Markov chain,\footnote{Although it may be thought of as a Markov chain Monte Carlo, our method is entirely unrelated to the Metropolis-Hastings algorithm.} and we may characterize it via the autocorrelation of various scalar quantities computed at each iteration.  We use the optimized global bandwidth $h$ (and adaptive sensitivity parameter $\alpha$, if not fixed to unity) for this purpose. 

After discarding a small number of initial iterations and then accumulating a number significantly greater than the autocorrelation time, we expect the collection of iterations to provide unbiased (though not necessarily independent) estimates of the population distribution.  For subsequent iterations we then use the median of $\hat{p}_{\rm pop}(\vec{\theta}_i^k)$ over a buffer of previous iterations (usually the previous 100) to determine the sample draw probabilities for the next iteration.  This population estimate should be more precise than one using only a single previous KDE;
in addition using the buffer estimate, the samples for each successive iteration are essentially independent. 

In reality, we do not expect the resulting population density estimate to be entirely unbiased, due to more fundamental limitations in both the \ac{PE} input~\citep[e.g.\ inaccuracies in the merger waveform model used, see][]{KAGRA:2021duu} and in the \ac{KDE} procedure itself.  Specifically, the expected bias of the \ac{KDE} is proportional to the second derivative of the true distribution function~\citep[][Section 3.3]{Silverman1986}, thus, depending on the eventual bandwidth choice, sharp peak or gap features may not be well represented.  This bias might be reduced by replacing the \ac{KDE} with a more general estimate, for instance Gaussian mixture models that allow the positions and weights of kernels to vary~\citep[e.g.][]{Rinaldi:2021bhm}, which though implies higher complexity and computational cost.

\subsection{One-dimensional mock data demonstration}

\begin{figure}[tbp]
    \centering
    \vspace{0.1cm}
    \includegraphics[width=0.7\linewidth, trim={0 0 0 0},clip]{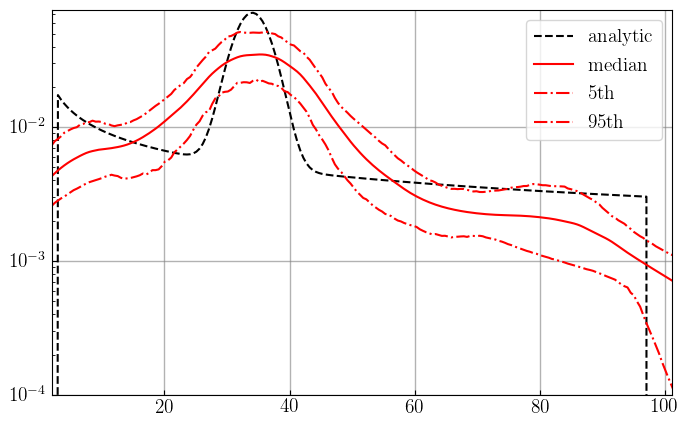}
    \\
    \vspace{0.2cm}
    \includegraphics[width=0.7\linewidth, trim={0 0 0 0},clip]{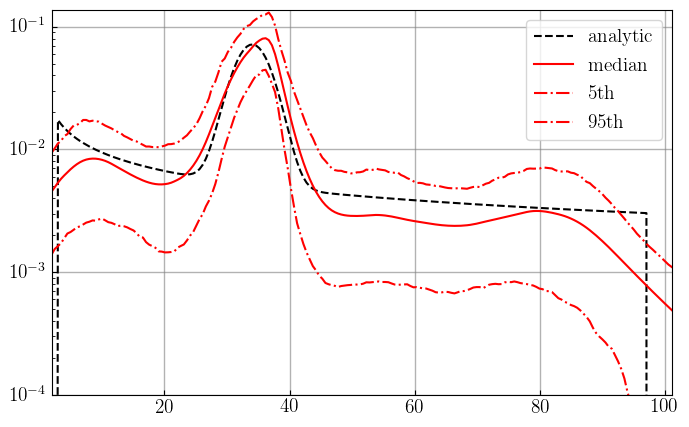}
    \caption{awKDE 
    for 60 events from a mock data mixture 
    distribution with 50\% power law ($\alpha = -0.5$) and 50\% Gaussian ($\mu=35$, $\sigma=3$) samples.
    100 mock parameter samples are generated for each event with measurement error $\sigma_m = 5$. 
    Top: awKDE drawing random parameter samples without reweighting. Bottom: applying iterative reweighting. The solid (dot-dashed) lines represent the median (symmetric 90\% confidence band) from 900 bootstrap iterations. 
    }
\label{fig:mockdatatest}
\end{figure}
We first test this iterative reweighting method on a simple mock dataset. 
We generate true event parameters for a mixture, drawing 30 events each from a truncated power law $p_{\rm pop}(x) \sim x^{-0.5}$ and from a Gaussian with mean (s.d.) of $\mu =35$ ($\sigma_p=3$) respectively. We then simulate ``measured'' event values with a Gaussian random scatter relative to the true parameters of s.d.\ $\sigma_m = 5$ (broader than the true Gaussian peak); for each measured value we generate 100 mock parameter samples with mean equal to the measured value and with the same uncertainty $\sigma_m = 5$. This procedure mimics \ac{PE} using an uninformative (flat) prior over $x$.

First, applying awKDE as in \citet{Sadiq:2021fin} to random draws from these mock parameter samples, as expected we find an over-dispersed estimate around the peak (Fig.~\ref{fig:mockdatatest}, top). 
Second, applying awKDE with our iterative reweighting algorithm we obtain the bottom plot of Fig.~\ref{fig:mockdatatest}: here the Gaussian height and s.d.\ appear accurately reconstructed and the true distribution is well within the 90\% percentiles of iteration samples, except near the step-function truncations of the power law which cannot be accurately represented by a Gaussian KDE. 
\begin{figure}[tbp]
    \centering
    \includegraphics[width=0.8\linewidth,trim={0 0 0 30},clip]{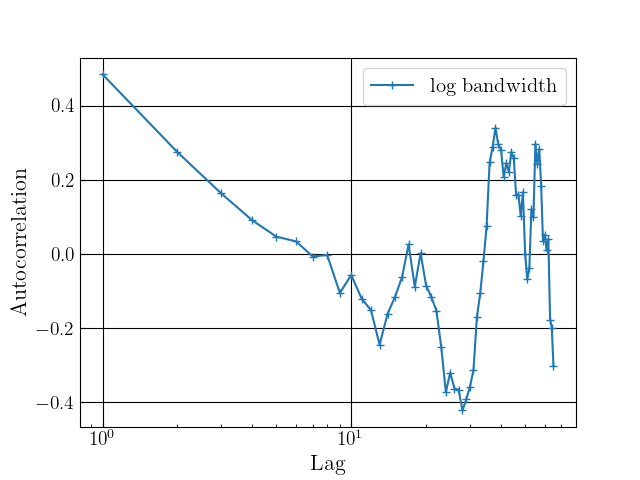}
    \caption{Autocorrelation of the (log) optimized global bandwidth series for the iterative Markov chain.  The x-axis shows lag, i.e.\ separation of the iterations between which autocorrelation is calculated.}
    \label{fig:onedmockdata_autocorr}
\end{figure}

We verify that the initial Markov process has accumulated several independent samples by plotting the autocorrelation of the optimized global bandwidth $h$ vs.\ lag (separation along the chain) in Fig.~\ref{fig:onedmockdata_autocorr}.\footnote{We fix the adaptive sensitivity parameter $\alpha$ to unity for 1-d data.}  The autocorrelation drops near zero by a lag of $\lesssim\!10$ iterations, thus a buffer of 100 iterations contains several independent population estimates.  (The estimate is more noisy for larger lags as fewer iterations are available.)

To further investigate possible statistical biases, we performed the same two awKDE analyses (unweighted and iterative reweighted) for 50 realizations of mock data with the same distribution. 
\begin{figure}[tbp]
\includegraphics[width=0.99\linewidth, trim={0 10 0 40},clip]{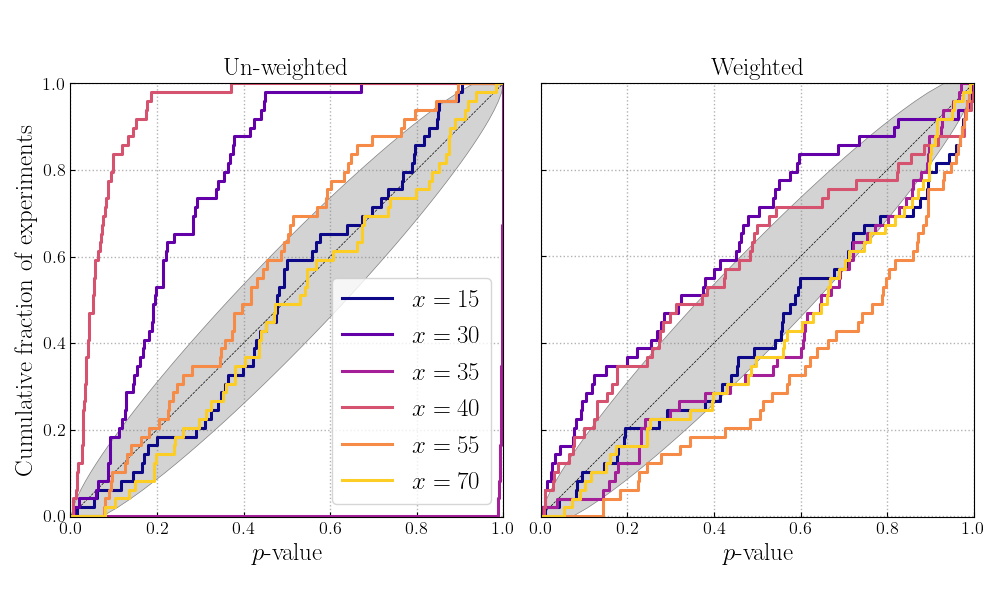}
    \caption{ P-P plots for 50 realizations of mock data as in Fig.~\ref{fig:mockdatatest}. 
    The x-axis is the percentile of the true population density relative to the KDE Monte Carlo samples, at each of several chosen parameter values. 
    The y-axis is the fraction of mock data experiments below the given percentile, while the gray area is an approximate 95 percentile confidence band.  Left: P-P plot for the unweighted awKDE.  Right: P-P plot for iterative reweighted awKDE. 
    Plots were created with  \url{https://lscsoft.docs.ligo.org/ligo.skymap/plot/pp.html}.
    }
\label{fig:pp-plot}
\end{figure}
At any given $x$ value, we can find the percentile ($p$-value) of the true $p_{\rm pop}$ relative to the awKDE Monte Carlo samples, and thus produce a P-P plot.  The first panel in Fig.~\ref{fig:pp-plot} for unweighted awKDE shows consistent estimates for values well away from the population peak, but catastrophic under- or overestimation close to the peak, as expected from strong over-dispersion.  For reweighted awKDE, the second panel shows much smaller biases close to the peak.  (We do not show $x$ values close to $0$ or $100$, as these are subject to large edge effects for both methods.)  While not entirely free of bias, the reweighted method offers much improved reconstruction of population features.

In the next subsection we verify the improved behaviour of the reweighted awKDE for two-dimensional mock data in the presence of correlated parameter uncertainties.

\subsection{Two-dimensional mock data test of iterative reweighting}
\label{app:2dmock}

\begin{figure}
\centering
    \includegraphics[width=0.95\linewidth]{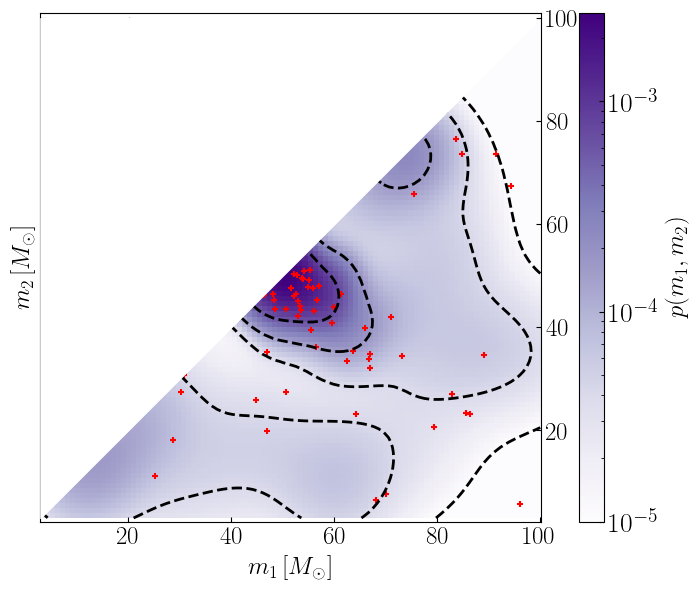}
    
    \includegraphics[width=0.95\linewidth]{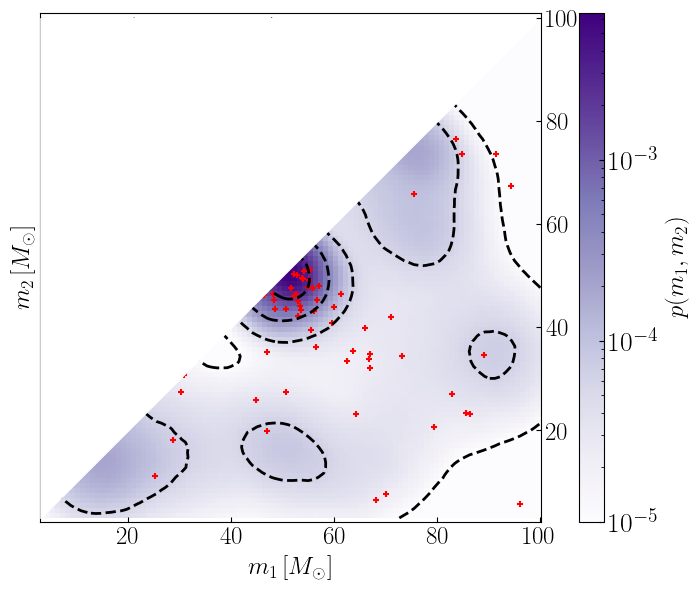}
    
    \caption{
    Density estimates for 2-d mock data: the true distribution is a mixture of a uniform distribution and a bivariate Gaussian with equal and uncorrelated variances in the two dimensions.  We simulate a measurement uncertainty which is strongly anti-correlated between the two dimensions to generate mock  parameter samples (see main text for details). 
    Top: unweighted awKDE from data with anti-correlated measurement uncertainties.  Red $+$ symbols show the measured event values while blue shading and contours show the estimated density. 
    The contours around (50, 50) are wider (elliptical) and clearly show an anti-correlation. Bottom: awKDE with iterative reweighting recovers the true distribution peak without (anti)correlation between parameters.}
\label{fig:mock2D}
\end{figure}
To investigate the performance of iterative reweighted KDE over 2-d data, we start with 60 mock events, 50\% from a two-dimensional uniform distribution and 50\% from a bi-variate normal (Gaussian) distribution. The normal distribution mean is $\mu = (50, 50)$ in arbitrary units and we take the two dimensions to be uncorrelated each with a variance of $9$. We then simulate measurement uncertainty in our mock data values with an anti-correlation between the two dimensions using covariance $\Sigma_p = \bigl(\begin{smallmatrix} 32.5 & -31.5\\-31.5 & 32.5 \end{smallmatrix}\bigr)$, corresponding to an error ellipse with 8:1 axes at $45^\circ$.  This is a simple choice to mimic the anti-correlation between $m_1$ and $m_2$ along a contour of constant chirp mass for lower-mass \ac{BBH}. As for the 1-d mock data, 100 parameter samples with the same covariance are generated around each measured value. 

We then investigate whether the reweighted awKDE can reconstruct the true (uncorrelated) population peak by reducing the artificial measurement uncertainty correlation.  The top panel of Fig.~\ref{fig:mock2D} for the unweighted awKDE (median over 900 iterations) clearly shows a biased reconstruction with anti-correlation around the peak at $(50, 50)$.\footnote{For 2-d data we impose the symmetry $f(m_2,m_1) = f(m_1,m_2)$ and hence only show the reconstructed density over half of the plane: this symmetry is discussed further in Sec.~\ref{ss:2d_mass_method}.} The bottom panel shows the results using the iterative reweighting scheme: the anti-correlation appears to be removed and the reconstructed peak is  as expected for the true distribution.

\begin{figure}
\centering
    \includegraphics[width=0.8\columnwidth, trim={0 0 0 5}, clip]{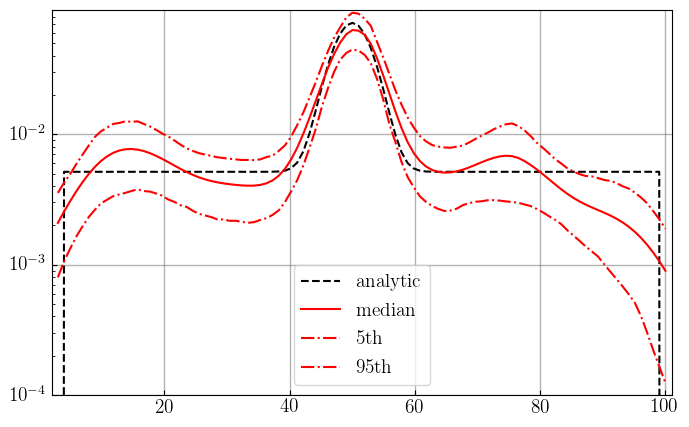}
    \caption{Integrated 1-d component distribution derived from 2-d mock data results. 
    The black dashed curve is the true 1-d distribution composed of uniform and Gaussian components with a Gaussian mean $\mu=50$. The red curves are derived from numerical integration of 2-d mock data results using iterative reweighted awKDE, as in Fig.~\ref{fig:mock2D}, bottom plot. 
    \label{fig:1dfrom2dmockdata}
}
\end{figure}

We also compute 1-d results from these 2-d  reweighted awKDE iteration samples by integrating over the second parameter (labelled $m_2$ in Fig.~\ref{fig:mock2D}) and compared with the corresponding 1-d true distribution. Fig.~\ref{fig:1dfrom2dmockdata} shows this comparison: the true Gaussian peak height and width are well recovered by the reweighted awKDE, although as in the 1-d case there are evident edge artifacts near the step-function truncations of the uniform distribution. 

\section{Results from GWTC-3}

We now apply the reweighted awKDE method to LVK observations. 
As in \cite{Sadiq:2021fin}, as input to our analysis we use parameter estimation samples~\citep{ligo_scientific_collaboration_and_virgo_2021_5546663} for the set of \ac{BBH} events catalogued in GWTC-3~\citep{LIGOScientific:2021djp,LIGOScientific:2023vdi} with false alarm rate below 1 per year.  For the sensitivity estimate we employ a fit to search injection (simulated signal) results~\citep{2019dan,sensitivityDan} released with the catalog~\citep{ligo_scientific_collaboration_and_virgo_2021_5636816}.

\subsection{One-dimensional mass distribution}
\begin{figure}[tbp]
    \includegraphics[width=\linewidth,trim={5 0 0 5},clip]{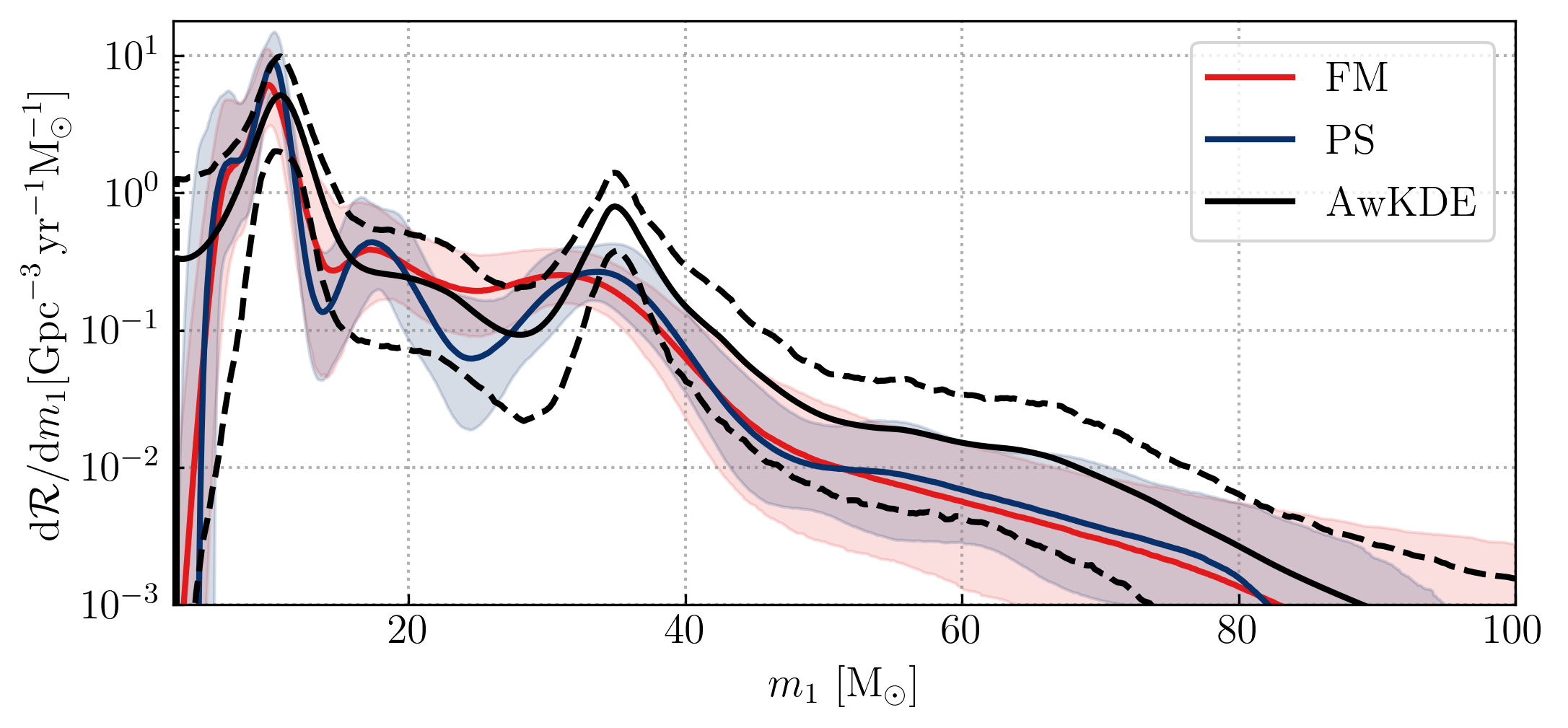}
    \caption{Differential rate over BBH primary mass using the iterative weighted KDE on GWTC-3 PE samples, assuming a power-law secondary distribution.  We overplot our estimates on two models, \textsc{Flexible mixtures} (FM), and \textsc{Power Law + Spline} (PS) in \citep{KAGRA:2021duu}.  Our estimate is generally consistent with other non-parametric methods, though with a higher and narrower peak around $35\,\msun$ and lacking any feature between the 10\,$\msun$ and 35\,$\msun$ peaks.
    }
    \label{fig:iterativerate_1d}
\end{figure}

We start by evaluating the effect of the iterative reweighting method on the 1-d primary mass distribution, taking 100 random PE samples for each of the 69 \ac{BBH} events; here, we assume a power-law distribution for secondary mass $p(m_2) \sim q^{1.5}$.  We reproduce the awKDE results from~\cite{Sadiq:2021fin} 
and use this estimate to seed the reweighted iteration algorithm.
%
After $\sim$1000 reweighting iterations 
we compute the median and symmetric 90\% interval from the last 900 rate estimates (the first 100 are used to set up a buffer for population weighting, as above): results are presented in Fig.~\ref{fig:iterativerate_1d}. 

Our estimate is generally consistent with other non-parametric or semi-parametric approaches, represented by the \textsc{Flexible mixtures}, and \textsc{Power Law + Spline} models in \cite{KAGRA:2021duu}, and does not show the over-dispersion apparent in Figure 8 of \cite{Sadiq:2021fin}; specifically, we find a slightly higher and narrower peak around $35\,\msun$, but no identifiable feature around $20\,\msun$ \citep[compare][]{Tiwari:2021yvr,Toubiana:2023egi}. 
%
%

\subsection{Two—dimensional Mass Reconstruction}
\label{ss:2d_mass_method}

Next, we apply our reweighting scheme on PE samples for both component masses and compute the two-dimensional (2-d) merger rate, using the estimated sensitive volume$\times$time (VT) as a function of the two masses.  We will first discuss various technical aspects of extending the 1-d calculation without assuming any power-law dependence for $m_2$. 

\paragraph{Binary exchange symmetry}
Typically when presenting binary parameter estimates, the convention $m_1>m_2$ is applied.  However, all aspects of binary formation physics and event detection and parameter estimation will be invariant under swapping the component labels, i.e.\ exchanging $m_1 \leftrightarrow m_2$ (at the same time exchanging the spins).  Thus, considering the differential merger rate $\mathcal R(m_1,m_2)$ as a function over the whole plane, it must have a reflection symmetry about the line $m_1=m_2$.  To respect this symmetry and remove biases resulting from the apparent lack of support at $m_2>m_1$, if using samples with the $m_1>m_2$ convention, we train and evaluate KDEs on \emph{reflected} sample sets, i.e.\ after adding 
copies of the samples with swapped components.  (Note also that a power-law $m_2$ distribution implies the rate is a non-differentiable function at the equal mass line, whereas a KDE by construction is smooth and differentiable everywhere.)

\paragraph{Choice of KDE parameters}
In \citet{Sadiq:2021fin}, we mainly considered a KDE constructed over linear mass (or distance) parameters; however, here we choose the logarithms of component masses.  While this choice is not expected to have a large impact on the results, since the kernel bandwidth is free to locally adapt in either case, it is technically preferable for a few reasons.  We avoid any possible KDE support at negative masses; there is a generally higher density of events towards lower masses considering the entire $3-100\,\msun$ range; the density of observed events also shows less overall variation over log coordinates; and when evaluating the KDE on a grid with equal spacing, fewer points are required to maintain precision for the low-mass region. 

For a 2-d KDE we also have a choice of kernel parameters, i.e.\ the Gaussian covariance matrix: given the similar or identical physical interpretation and range of values between $\ln m_1$ and $\ln m_2$, we choose a covariance proportional to the unit matrix, with an overall factor determined by the local adaptive bandwidth for each event. 

\paragraph{PE prior}
The \ac{PE} samples released by LVK use a prior uniform in component masses~\citep{ligo_scientific_collaboration_and_virgo_2021_5546663} up to a factor dependent on cosmological redshift; we currently do not consider reweighting relative to the default cosmological model. 
As the prior is a density, it transforms with a Jacobian factor when changing variables to $\ln m_1, \ln m_2$, thus, we must divide the estimated rate $\mathcal{R}(\ln m_1, \ln m_2)$ by a prior $\propto m_1 m_2$ when obtaining reweighted draw probabilities. 
 \begin{figure}[tbp]
        \includegraphics[width=0.99\columnwidth,trim={2 0 2 3},clip]{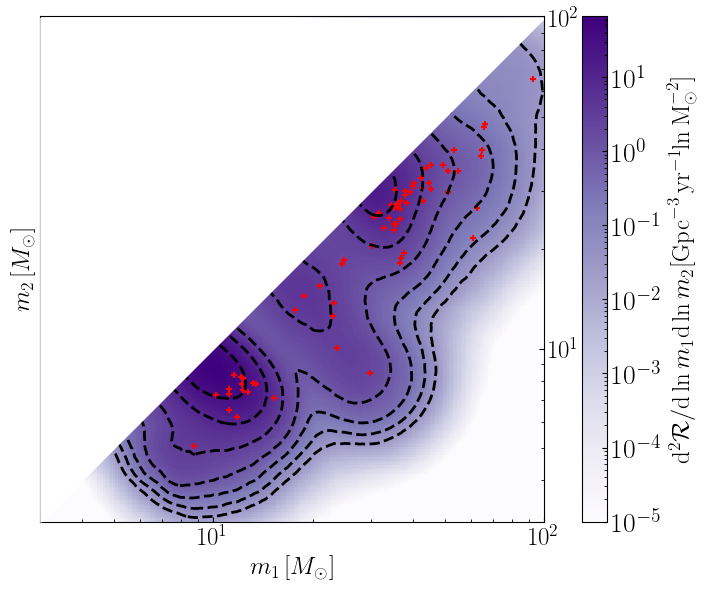}
        \caption{Merger rate over binary component masses, estimated using 69 BBH events from GWTC-3.  Red $+$ symbols show the median mass samples for each event. The contours and color scale show the two-dimensional differential merger rate over $\ln m_1, \ln m_2$ from iterative reweighted KDE (median over 900 iterations). 
        Two main maxima are visible at $m_1\,(m_2)\sim 10\,(8)\,\msun$ and $\sim\! 35\,(32)\,\msun$ with a possible less significant overdensity around $m_1\sim 20\,\msun$.}
    \label{fig:twod_iterativerate}
\end{figure}

With these technical choices, we perform 1500 reweighting iterations in total, the first 600 using the Markov chain (i.e.\ the immediately preceding rate estimate) for sample draw weights, and the remaining 900 using the buffer median estimate. 
\begin{figure}[tbp]
\centering 
\includegraphics[width=0.8\linewidth,trim={0 0 0 25},clip]{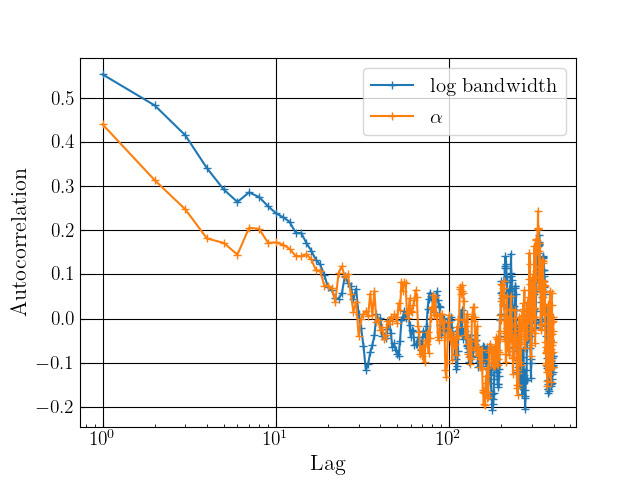}
    \caption{Autocorrelations of the KDE (log) global bandwidth and adaptive parameter $\alpha$ using the initial 600 Markov chain iterations.  The autocorrelation drops close to 0 by a lag of $\sim$30 iterations.  The estimate becomes noisy at high lags as a smaller number of iterations is available.
    \label{fig:correlationbwd}
}
\end{figure}
Fig.~\ref{fig:twod_iterativerate} shows the rate estimate computed with iterative reweighting for BBH events in GWTC-3 (median over the last 900 iterations). 
The autocorrelations of optimal global bandwidth and sensitivity parameter $\alpha$ for the first 600 iterations are shown in Fig.~\ref{fig:correlationbwd}: the correlation drops close to $0$ at a lag of $\sim\,$30 iterations. 

The mass distribution shows several interesting features in addition to the expected peaks (overdensities) around primary masses of $\sim\!10\,\msun$ and $\sim\!35\,\msun$, with corresponding peaks over secondary mass. 
For primary masses $\sim\!35\,\msun$ up to $80\,\msun$, the most likely secondary mass is 
$\sim\!30\,$--$\,35\,\msun$.  Thus, over this range the two component masses appear almost independently chosen.  Around the $m_1 \sim 10\,\msun$ peak, some \emph{anti-correlation} of the two components appears, i.e.\ higher $m_1$ favors lower $m_2$.  We investigate the significance of this feature by calculating the median of the $m_2$ distribution as a function of primary mass, i.e.\ $m_{2}^{50\%}(m_1)$; this is a decreasing function over a range of $m_1$ from $\sim$$10$ through $\sim$$14\,\msun$ for $90$--$95\%$ of our sample estimates, depending on the exact range of $m_1$ chosen. The anti-correlation is thus moderately but not highly significant.

Between the two peaks the distribution of mass ratios appears broader than at either one \citep[as hinted at in][]{Tiwari:2023xff}, although the apparent trend is based on a small number of events.  We also see a narrow lower density region just above the $\sim\!10\,\msun$ peak (cf.\ the local minimum at chirp mass $\sim\!11\,\msun$ in \citealt{Tiwari:2023xff}). 

\begin{figure}[tbp]
    \includegraphics[width=\linewidth,trim={0 0 0 3},clip]{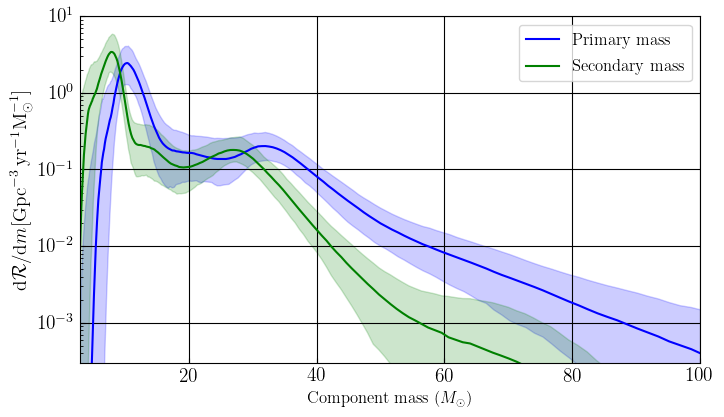}
    \caption{Merger rates over component masses, obtained by numerical integration of our 2-d KDE rate estimates in Fig.~\ref{fig:twod_iterativerate}: the medians and symmetric 90\% confidence regions are shown.}
    \label{fig:m1m21Drates}
\end{figure}
We also integrated the 2-d KDE rate estimate numerically over both $m_1$ and $m_2$ to obtain merger rates over component mass.  As shown in Fig.~\ref{fig:m1m21Drates}, we recover features consistent with the 2-d estimates and with other methods.  Each component mass distribution appears well modelled by a combination of two Gaussian peaks and (broken) power laws. 

%
%

\begin{figure}[tbp]
\centering
    \includegraphics[width=0.85\linewidth]{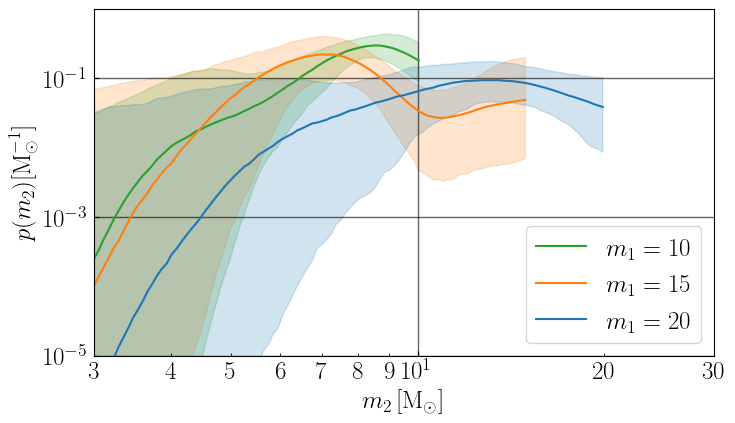} \\
    \includegraphics[width=0.85\linewidth]{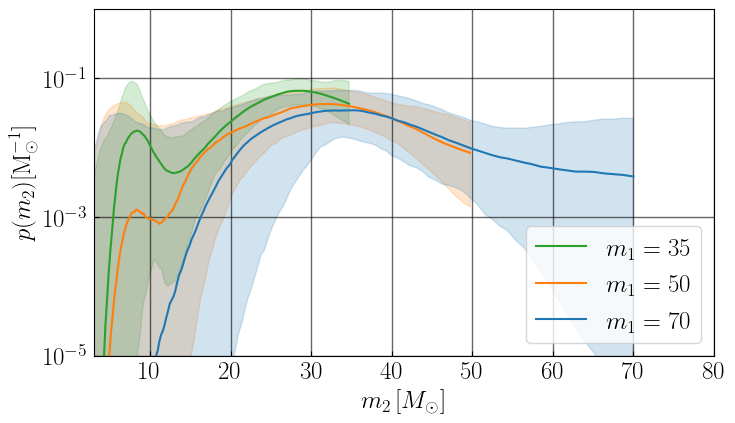}
    \caption{Secondary mass $m_2$ distributions estimated via iterative reweighted KDE for various fixed values of primary mass: for each $m_1$ (in $\msun$) we plot the median and symmetric $90\%$ confidence band.  To better visualize features in the distributions we use a log (linear) $m_2$ scale in the upper (lower) plot respectively. 
    }
\label{fig:secondarymassfixedm1}
\end{figure}
To elucidate features in the 2-d distribution, we choose various representative values of primary mass to plot the distribution of $m_2$ in Fig.~\ref{fig:secondarymassfixedm1}.  The similarity between secondary distributions for $m_1 \gtrsim 35\,\msun$ is evident.

\begin{figure}[tbp]
\centering
    \includegraphics[width=0.85\linewidth,trim={0 0 0 3},clip]{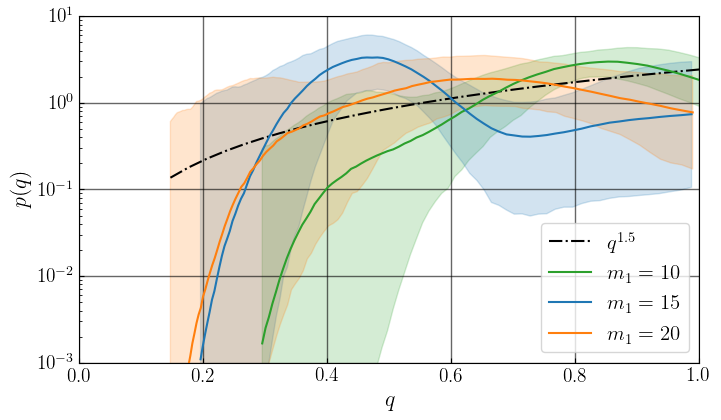} \\
    \includegraphics[width=0.85\linewidth]{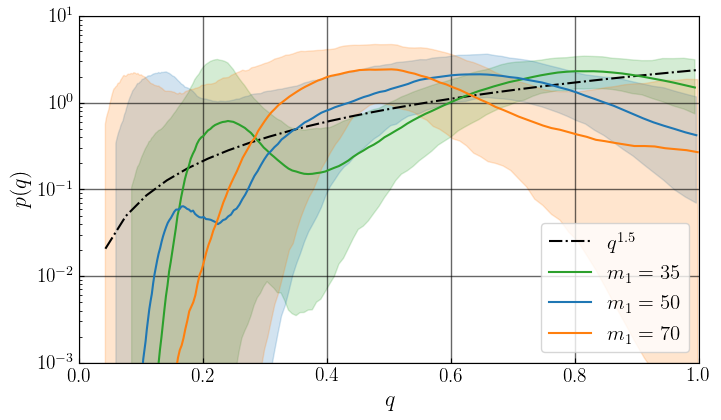}
    \caption{Mass ratio $q$ distributions estimated via iterative reweighted KDE for various fixed values of primary mass: for each $m_1$ (in $\msun$) we plot the median and symmetric $90\%$ confidence band. For comparison we overplot a power law dependence $\propto q^{1.5}$.
    }
\label{fig:massratio}
\end{figure}
We may also derive the distribution of mass ratio $q$ from our 2-d rate estimate. 
We plot this for various representative values of primary mass in Fig.~\ref{fig:massratio}, and compare to a typical power law $\propto q^{1.5}$.  First, we see that the $q$ distribution varies over primary mass; hence, models where it is forced to the same form over the whole mass range are likely to have nontrivial bias.  For some primary masses, $p(q)$ is consistent with a monotonic increasing function such as a positive power, but for others it clearly decreases over some of the range.  Roughly, if $m_1$ is close to a peak then $p(q)$ is consistent with an increasing power law, but for other values the mass ratio rather shows a maximum at intermediate values, down to $q\sim 0.5$ for $m_1=15$ or $m_1=70$.

This behaviour may suggest that the primary and secondary masses are independently drawn from similar distributions, modulo a $q$-dependent ``pairing factor''~\citep{Fishbach:2019bbm} which influences the relative probability of binary merger; (recent work by \citet{Farah:2023swu} further explores the case of similar or identical underlying primary and secondary mass distributions).  In any case, the preference towards $q\sim 1$ seen in previous work is not confirmed here.  \cite{Callister:2023tgi} reached a similar conclusion, although assuming a ``universal'' distribution $p(q)$ over all primary masses.

\paragraph{Results including GW190814}
We also estimate the 2-d and 1-d integrated merger rates using our iterative reweighted KDE method when the outlier event GW190814~\citep{LIGOScientific:2020zkf}, which has a mass ratio $\mathcal{O}(10)$ and a secondary mass barely above the likely neutron star maximum mass, is included in the BBH population.  Detailed results are presented in Appendix~\ref{app:190814}: roughly summarizing the trends seen there, the bulk of the estimated distribution remains little changed by the addition of the extra event, although the peak in secondary mass below $10\,\msun$ is shifted towards lower values and both higher and broader; this is likely due to a general increase in KDE bandwidth when optimized with cross-validation.  (In parameterized models  the estimated mass distribution is also highly sensitive to inclusion of GW190814~\citep{Abbott:2020gyp,KAGRA:2021duu}.)  It is not clear whether other methods for bandwidth choice would yield more accurate estimates; higher event statistics in the low $m_2$ regime are clearly desirable. 


\section{Discussion}

\paragraph{Summary of results}
In this work we undertook a detailed investigation of the full 2-d mass distribution of merging compact binary black holes observed by LIGO-Virgo-KAGRA up to the O3 run without assuming any specific functional form for the secondary mass or mass ratio, enabled by a new method of iterative density estimation to address mass measurement uncertainties.  Although we reproduce the broad features and local maxima seen in other parametric and non-parametric analyses, we find significantly less preference for near-equal masses than in most previous works; we also find that the mass ratio distribution cannot be described by a single function over the whole population \citep[compare][]{Tiwari:2023xff}.  For a range of primary masses, we find non-monotonically varying secondary and mass ratio distributions, thus a power-law dependence is ruled out. 
Furthermore, we find that for primary masses above $35\,\msun$ the secondary mass distribution is nearly independent of $m_1$, with a ``preferred partner'' mass of $m_2 \simeq 30-35\,\msun$.  Conversely, near the low-mass peak $m_1\simeq 10\,\msun$ we observe an anti-correlation between the two components, i.e.\ higher $m_1$ implies lower $m_2$. 

\paragraph{Possible astrophysical interpretations}
Our new estimate of the joint $m_1$--$m_2$ distribution may be compared to model predictions in the literature; because our individual component marginal distributions are similar to previous findings, we focus here on the mass ratio. Broadly speaking, we can distinguish model predictions from the isolated binary and dynamical channels.

The isolated binary channel predicts relatively flat $p(q)$ distributions \citep[e.g.][]{2020A&A...636A.104B,2021A&A...651A.100O} compared to the dynamical channel (see Fig. 2 in \citealt{Baibhav:2022qxm} and Fig.~1 in \citealt{Zevin:2020gbd}). Our estimates show in general flatter $q$ distributions than the GWTC-3 results presented in \citet{KAGRA:2021duu}. 

Because predictions for $p(q)$ in the isolated binary channel depend strongly on the adopted parameters \citep[see e.g.][]{2022MNRAS.516.5737B}, our results provide an important step towards constraining astrophysical parameters with GWs.  For example, the steep $p(q)$ found for very small common envelope efficiency parameter \citep[$\alpha_{\rm CE} \simeq 0.2$,][]{Baibhav:2022qxm} and the chemically homogeneous evolution model \citep{2016MNRAS.458.2634M} seem disfavoured, implying that these routes cannot account for the majority of the observed population. 

The stable mass transfer channel is efficient for primary masses near $\sim\!10\,\msun$. \cite{2022ApJ...940..184V} predicts a dearth of near-equal mass mergers, which is because the binary needs to be relatively unequal in mass during the second mass transfer phase for the orbit to shrink, but not too unequal to avoid unstable mass transfer.  This is only partly supported by our Fig.~\ref{fig:twod_iterativerate} in that the low-mass peak has support from equal mass out to $q\simeq 0.5$.  For some parameter choices their models predict bi-modality in $p(q)$, with peaks at $q\simeq0.35$ and $q\simeq0.75$: our results suggest a peak at $q\simeq0.8$ for $m_1\simeq10\,\msun$ and at $q\simeq0.45$ for $m_1\simeq15\,\msun$ (see Fig.~\ref{fig:massratio}), suggesting that a more detailed comparison may yield interesting constraints.

For the dynamical channel, it is interesting to consider whether models now predict $q$ distributions that are too steeply rising.  \citet{2016PhRvD..93h4029R} modelled BBH mergers that formed dynamically in globular clusters: they find a median mass ratio of 0.87, with 68\% of sources having mass ratios $q>0.8$.  As shown in Fig.~\ref{fig:massratio} we find comparable support for near-equal mass only at $m_1 \sim 10\,\msun$ or $\sim\!35\,\msun$;  elsewhere our median $q$ is significantly lower.  

\citet{2023MNRAS.522..466A} model BBH merger in globular clusters in comparison to the GWTC-3 $q$ distribution: their model distributions are flatter and underestimate the power-law LVK fits by an order of magnitude at $q\simeq1$. 
They find a final $q$ distribution flatter than the $q$ distribution of dynamically formed BBHs ($p(q)\propto q^4$ for metal-poor clusters) because the BH mass function is not always sufficiently sampled, such that a secondary BH with a mass similar to the primary is present in a cluster; and due to a slight bias against equal-mass BBH due to their lower inspiral probability. 
The reported $p(q)$ in their Fig.~1 is relatively flat for $q\gtrsim0.7$, qualitatively in agreement with our findings for $m_1\gtrsim 20\,\msun$ (Fig.~\ref{fig:massratio}, lower panel; their models cannot reproduce observed rates for lower-mass primaries.)  Due to the predicted pair instability gap, all BBH mergers in their models with $m_1\gtrsim50\,\msun$  are hierarchical mergers, i.e.\ a BBH in which at least one of the components is a BBH merger remnants that was retained in the cluster \citep[e.g.][]{Antonini2016,Rodriguez:2019huv,Kimball:2020qyd}.  Mergers with second-generation primaries are expected to have a mass ratio $q\simeq 0.5$, which is supported  by our distribution for $m_1=70\,\msun$ (Fig.~\ref{fig:massratio}). 

The $p(q)$ distribution is expected to be slightly flatter for dynamically formed BBHs in young (open) star clusters, because they have fewer BHs per cluster and their higher metallicities lead to steeper BH mass functions and therefore lower companion masses \citep[e.g.][]{2021MNRAS.503.3371B}. 
A accurate picture of the mass ratio distribution is therefore important to understand the relative contribution of dynamically formed BBHs in young (and metal-rich) and old (and metal-poor) star clusters; and also more generally for understanding the relative contributions of isolated and dynamically formed binaries in the population as a whole \citep{Zevin:2020gbd,Baibhav:2022qxm}. 

An intriguing apparent feature in our reconstruction, the anti-correlation between $m_1$ and $m_2$ in the low-mass ($m\sim 10\,\msun$) peak, suggests a connection to isolated binary dynamics, though it would be premature to link it with a specific mechanism.

\paragraph{Technical issues and biases}
As noted in the introduction, measurement errors of the binary mass ratio are correlated with those in (orbit-aligned) spins.  Since we have so far not attempted to reconstruct or estimate the merging binary spin distribution, we implicitly assume that distribution is equal to the prior used for parameter estimation (uniform in magnitude and isotropic in direction): this is a potential source of bias which remains to be addressed by future work. The distribution of aligned spins has been found to be concentrated near zero, with a slight preference for positive aligned spin~\citep{Miller:2020zox,Abbott:2020gyp,KAGRA:2021duu}; hence, the degree of bias may be limited.  \cite{Callister:2021fpo} also note the intriguing possibility that the \emph{true} mass ratio and aligned spin (after allowing for measurement errors) are anti-correlated. 

A converse question concerns inferences on BH spin distributions which either assume a specific distribution in $q$, or a power law with index as a hyperparameter: if the $p(q)$ model is significantly inaccurate, are such spin inferences biased? (\citealt{Ng:2018neg} and~\citealt{Miller:2020zox} contain detailed discussion of potential biases in aligned spin population estimates.)  The effect may not be large, as most \ac{BBH} events by necessity have parameter values close to the observed peaks, for which we find a $q$ distribution which is not far from power-law. 

\paragraph{Extensions of the method}
As already noted, here we restricted the application of our KDE to the binary mass distribution; component spins, and distance or redshift are then the next relevant parameters for population analysis.  We expect to encounter a technical issue in optimizing the Gaussian kernel for a multi-dimensional data set, where it will not be appropriate (or even meaningful, given the different units) to impose equal variances over different parameters as we currently do for (log) $m_1$ and $m_2$.  For more than two dimensions a grid search may not be practicable; more sophisticated methods may be required in order to realize the potential of iterative KDE over a full set of population parameters.

\section*{Acknowledgements}
We thank Daniel Wysocki for making available fitted sensitivity estimates for binary mergers in the O1-O3 data. 
We also benefited from conversations with Lieke van Son, Floor Broekgaarden and Fabio Antonini, and with Will Farr, Thomas Callister, Amanda Farah, Vaibhav Tiwari and others in the LVK Binary Rates \& Populations group. 
This work has received financial support from Xunta de Galicia (CIGUS Network of research centers), by European Union ERDF 
and by the ``María de Maeztu'' Units of Excellence program CEX2020-001035-M and the Spanish Research State Agency.  TD and JS are supported by research grant PID2020-118635GB-I00 from the Spanish Ministerio de Ciencia e Innovaci{\'o}n.  JS also acknowledges support from the European Union’s H2020 ERC Consolidator Grant ``GRavity from Astrophysical to Microscopic Scales'' (GRAMS-815673) and the EU Horizon 2020 Research and Innovation Programme under the Marie Sklodowska-Curie Grant Agreement No.\ 101007855. MG acknowledges support from the Ministry of Science and Innovation (EUR2020-112157, PID2021-125485NB-C22, CEX2019-000918-M funded by MCIN/AEI/10.13039/501100011033) and from AGAUR (SGR-2021-01069). 

The authors are grateful for computational resources provided by the LIGO Laboratory and supported by National Science Foundation Grants PHY-0757058 and PHY-0823459. This research has made use of data or software obtained from the Gravitational Wave Open Science Center (gwosc.org), a service of the LIGO Scientific Collaboration, the Virgo Collaboration, and KAGRA. This material is based upon work supported by NSF's LIGO Laboratory which is a major facility fully funded by the National Science Foundation, as well as the Science and Technology Facilities Council (STFC) of the United Kingdom, the Max-Planck-Society (MPS), and the State of Niedersachsen/Germany for support of the construction of Advanced LIGO and construction and operation of the GEO600 detector. Additional support for Advanced LIGO was provided by the Australian Research Council. Virgo is funded, through the European Gravitational Observatory (EGO), by the French Centre National de Recherche Scientifique (CNRS), the Italian Istituto Nazionale di Fisica Nucleare (INFN) and the Dutch Nikhef, with contributions by institutions from Belgium, Germany, Greece, Hungary, Ireland, Japan, Monaco, Poland, Portugal, Spain. KAGRA is supported by Ministry of Education, Culture, Sports, Science and Technology (MEXT), Japan Society for the Promotion of Science (JSPS) in Japan; National Research Foundation (NRF) and Ministry of Science and ICT (MSIT) in Korea; Academia Sinica (AS) and National Science and Technology Council (NSTC) in Taiwan.

\bibliography{reference}

\appendix

\section{Results including GW190814}
\label{app:190814}

\begin{figure}[tbp] 
    \centering
    \includegraphics[width=0.45\columnwidth, trim={10 0 0 0},clip]{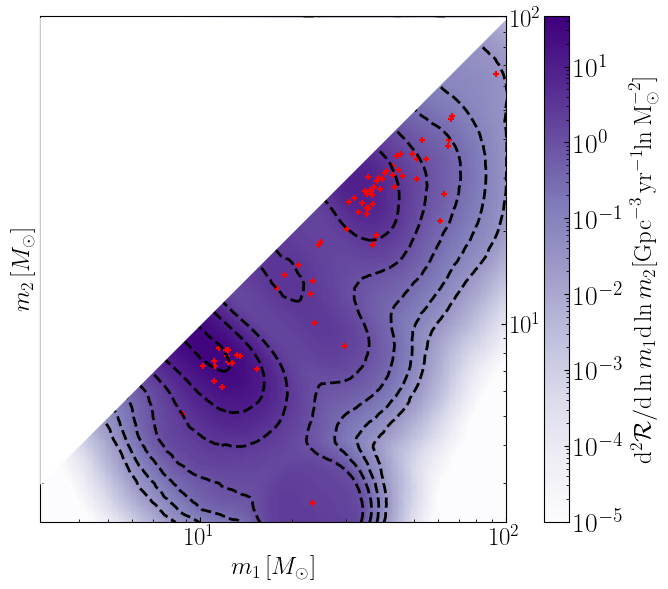}
    \hspace{0.5cm}
    \includegraphics[width=0.5\columnwidth]{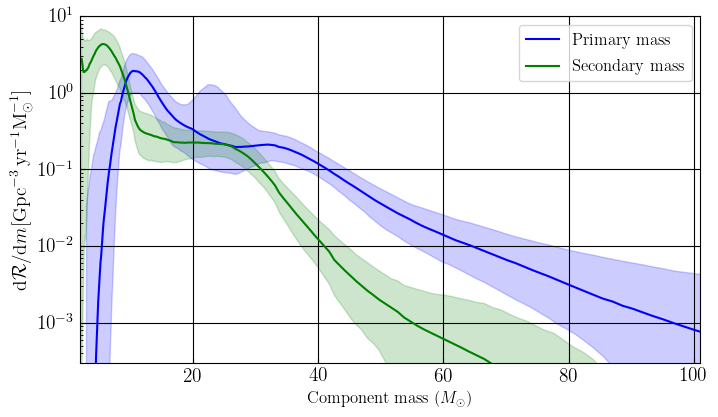}
    \caption{Left: Two dimensional merger rate density estimated via reweighted awKDE with for 70 events (69 \ac{BBH} plus GW190814) from GWTC-3.  Red $+$ symbols show the median mass samples for each event. 
    Right: The corresponding one dimensional component mass distributions from numerical integration of 2-d results.}
    \label{fig:twod_iterativeratewithgw190814}
\end{figure}
In addition to our main results from significant \ac{BBH} events in GWTC-3, we examine the effect of including the outlier event GW190814 \citep{LIGOScientific:2020zkf} with low secondary mass $m_2= 2.59\,\msun$ $(q=0.11)$,  which may be consistent either with a very massive neutron star or light black hole.  We slightly extend the range of $m_2$ over which the KDE is evaluated in order to include the additional event.  Fig.~\ref{fig:twod_iterativeratewithgw190814} summarizes our results, which are also briefly discussed in the main text. 

\end{document}